\documentclass[12pt]{iopart}

\usepackage{graphicx}
\def\Journal#1#2#3#4{ #4 {#1} {\bf#2} #3}

\def\NPA{{\em Nucl. Phys.} A}
\def\NPB{{\em Nucl. Phys.} B}
\def\PLB{{\em Phys. Lett.}  B}
\def\PRL{\em Phys. Rev. Lett.}
\def\PRD{{\em Phys. Rev.} D}
\def\PRC{{\em Phys. Rev.} C}

\def\JPG{{\em J. Phys.} G}

\def\RMP{\em Rev. Mod. Phys.}
\def\IJMPA{{\em Int. J. Mod. Phys.} A}
\def\JHEP{\em J. High Energy Phys.}
\def\PTP{\em Prog. Theor. Phys.}

\def\ep{\epsilon}
\def\lam{\lambda}

\def\la{\langle}
\def\ra{\rangle}

\def\al{\alpha}

\def\be{\begin{equation}}
\def\ee{\end{equation}}
\def\bea{\begin{eqnarray}}
\def\eea{\end{eqnarray}}

\begin{document}

\title{Exclusive rare
$B_s\to (K,\eta,\eta')\ell^+\ell^-$ decays in the light-front quark model}

\author{Ho-Meoyng Choi }

\address{
Department of Physics, Teachers College, Kyungpook National
University, Daegu, Korea 702-701}
\ead{homyoung@knu.ac.kr}
\begin{abstract}
Using the light-front quark model, we calculate the transition form factors,
decay rates, and longitudinal lepton polarization asymmetries for
the exclusive rare $B_s\to (K,\eta^{(\prime)})(\ell^+\ell^-,\nu_{\ell}\bar{\nu_{\ell}}$
($\ell=e,\mu,\tau$) decays within the standard model, taking into account the $\eta-\eta'$
mixing angle.
For the mixing angle $\theta=-20^{\circ}$ ($-10^{\circ}$) in the octet-singlet basis,
we obtain
${\rm BR}(B_s\to \eta\sum\nu_{\ell}\bar{\nu}_{\ell})=1.1\; (1.7)\times 10^{-6}$,
${\rm BR}(B_s\to \eta\mu^+\mu^-)=1.5 \;(2.4)\times 10^{-7}$,
${\rm BR}(B_s\to \eta\tau^+\tau^-)=3.8 \;(5.8)\times 10^{-8}$,
${\rm BR}(B_s\to \eta'\sum\nu_{\ell}\bar{\nu}_{\ell})=1.8\; (1.3)\times 10^{-6}$,
${\rm BR}(B_s\to \eta'\mu^+\mu^-)=2.4 \;(1.8)\times 10^{-7}$, and
${\rm BR}(B_s\to \eta'\tau^+\tau^-)=3.4 \;(2.6)\times 10^{-8}$, respectively.
The branching ratios for
the $B_s\to K(\nu_{\ell}\bar{\nu_{\ell}},\ell^+\ell^-)$ decays are at least an order of
magnitude smaller than
those for the $B_s\to \eta^{(\prime)}(\nu_{\ell}\bar{\nu_{\ell}},\ell^+\ell^-)$ decays.
The averaged values of the lepton polarization asymmetries
for $B_s\to (K,\eta^{(\prime)})\ell^+\ell^-$
are obtained as
$\la P^K_L\ra_\mu=\la P^\eta_L\ra_\mu=\la P^{\eta'}_L\ra_\mu=-0.98$, $\la P^K_L\ra_\tau=-0.24$,
 $\la P^\eta_L\ra_\tau=-0.20$ and $\la P^{\eta'}_L\ra_\tau=-0.14$,
respectively.
\end{abstract}

\maketitle
\section{Introduction }
The study of the exclusive decays in the beauty sector
allows one to explore the standard model (SM) and search for new physics  effects.
The B factory experiments such as BaBar at SLAC, Belle at KEK, LHCb at CERN, and
B-TeV at Fermilab make precision tests of the SM and beyond the SM ever more promising.
Especially, the $B_s$-meson system becomes a key element in the $B$-physics program of B factories
ever since the first evidence for $B_s$ production at the $\Upsilon(5S)$ was found by
the CLEO collaboration~\cite{CLEO05,CLEO06}.  The D0~\cite{D008} and CDF~\cite{CDF08} Collaborations
have made measurements of the charge-parity (CP) violating weak $B_s-\bar{B}_s$
mixing phase $\phi_s$ in $B_s\to J/\psi\phi$ decays.
While the SM expectation $\phi^{\rm SM}_s$~\cite{Lenz,Bona} is nearly zero,
the measured $\phi_s$ differs from 0 by more than $3\sigma$ (but with a sizable error).
This measurement of $\phi_s$ inconsistent
with zero (if confirmed) would indicate
an evidence of new physics. Recently, the Belle Collaboration also
measured the branching ratios of the $B_s\to J/\psi\phi$ and $B_s\to J/\psi\eta$ decays
and the preliminary result~\cite{Dr} of the $B_s\to J/\psi\phi$ decay is about 3 times larger than
that for the $B_s\to J/\psi\eta$ decay.
This ratio agrees with a rough estimate obtained within the naive
quark model (neglect octet-singlet mixing),
where the $s\bar{s}$ part of the $\eta$ meson wave function is one third in contrast
to the fully $s\bar{s}$ content of $\phi$ mesons.
With the upcoming chances that a numerous number of $B_s$ mesons will be produced
at hadron colliders, one might explore the exclusive rare $B_s$ decays to
$(K,\eta,\eta')\ell^+\ell^-$(and $\nu_{\ell}\bar{\nu}_{\ell}$)  ($\ell=e,\mu,\tau$)
induced by the flavor-changing neutral current (FCNC) transitions
$b\to(d,s)$. Since in the SM the rare $B_s$ decays are
forbidden at tree level and occur at the lowest order only through one-loop penguin
diagrams~\cite{GWS,BM,Misiak,TI,BBL,AMM,KMS,AKS}, the rare $B_s$ decays are well suited to test the SM and detect
new physics effects.  While the experimental tests of exclusive decays are much easier than
those of inclusive ones, the theoretical understanding of exlcusive
decays is complicated mainly due to the nonperturbative hadronic
form factors entered in the long distance nonperturbative contributions.
Therefore, a reliable estimate of the hadronic form factors
for the exclusive rare $B_s$ decays is very important to make correct predictions within
and beyond the SM. The $\eta-\eta'$ mixing angle may also be extracted from the
rare $B_s$ decays to $\eta$ and $\eta'$ final states.

In our previous work~\cite{CJK}, we have analyzed the exclusive rare
$B\to K\ell^+\ell^-$ decays
within the framework of the SM, using our light-front
quark model (LFQM) based on the QCD-motivated effective LF Hamiltonian~\cite{CJ1,CJ_PLB1,JC_E}.
The experimental values of the branching ratios
${\rm BR}(B\to K\ell^+\ell^-)=(0.75^{+0.25}_{-0.21}\pm 0.09)\times 10^{-6}$
from Belle~\cite{Abe02} and
$(0.34\pm 0.07\pm 0.02)\times 10^{-6}$ ($\ell=e,\mu$)
from BABAR~\cite{Aubert06} detectors are consistent with our LFQM prediction
$0.5\times 10^{-6}$~\cite{CJK}
based on the SM. Recently, we also analyzed $B_c$ properties and various exclusive decay
modes  such as the semileptonic
$B_c\to (D,\eta_c,B,B_s)\ell\nu_\ell$ decays~\cite{CJBc}, the
rare $B_c\to D_{(s)}\ell^+\ell^-$~\cite{Bcrare},
and the nonleptonic two-body $B_c\to (D_{(s)},\eta_c,B_{(s)})(P,V)$ decays~\cite{CJNRD}
(here $P$ and $V$ denote pseudoscalar and vector mesons, respectively).
The form factors $f_{\pm}(q^2)$ and $f_T(q^2)$ for the exclusive rare decays~\cite{Bcrare}
between two pseudoscalar mesons are obtained in the Drell-Yan-West ($q^+=q^0+q^3=0$)
frame~\cite{DYW} (i.e., $q^2=-{\bf q}^2_\perp<0$), which is useful because only the
valence contributions are needed unless the zero-mode contribution exists.
The covariance (i.e., frame independence) of our model has been checked by performing the
LF calculation in the $q^+=0$ frame in parallel with the manifestly
covariant calculation using the exactly solvable covariant fermion
field theory model in $(3+1)$ dimensions. We also found the zero-mode contribution
to the form factor $f_-(q^2)$ and identified~\cite{CJBc} the zero-mode operator that
is convoluted with the initial and final state LF wave functions.

The purpose of this paper is to extend our our
LFQM~\cite{CJK,CJ1,CJ_PLB1,JC_E,CJBc,Bcrare,CJNRD} to calculate the hadronic form factors,
decay rates and the longitudinal lepton polarization asymmetries (LPAs) for the exclusive rare
$B_s\to (K,\eta, \eta')\ell^+\ell^-$
and $\nu_{\ell}\bar{\nu}_{\ell}$ decays within the SM.
The LPA, as another parity-violating
observable, is an important asymmetry~\cite{Hew} and could be
measured at hadron colliders such as LHCb.
In particular, the $\tau$ channel would be more accessible
experimentally than $e$- or $\mu$-channels since the LPAs
in the SM are known to be proportional to the lepton
mass.
There are some theoretical approaches to the calculations of the exclusive rare
$B_s\to \eta\ell^+\ell^-$~\cite{Skands,Geng,CCD} and
$B_s\to \eta'\ell^+\ell^-$~\cite{Geng,CCD} decays, but not the
$B_s\to K\ell^+\ell^-$ decay mode as far as we know.

The paper is organized as follows. In Sec. 2, the SM operator basis,
describing the $b\to (d,s)(\ell^+\ell^-,\nu_{\ell}\bar{\nu}_\ell)$
transitions, is presented.
In Sec. 3, we briefly describe the formulation of our LFQM and the procedure
of fixing the model parameters using the variational principle for the QCD
motivated effective Hamiltonian.
We discuss the  rare decays between two pseudoscalar mesons using an exactly
solvable model based on the covariant Bethe-Salpeter (BS) model of $(3+1)$-dimensional
fermion field theory and show the equivalence between the
results obtained by the manifestly covariant method and the LF method
in the $q^+=0$ frame. We then present the
LF covariant forms of the form factors $f_{\pm}(q^2)$ and
$f_T(q^2)$ obtained from our LFQM. The $\eta-\eta'$ mixing angle for the
$B_s\to\eta^{(\prime)}$ transitions is also discussed in this section.
In Sec. 4, our numerical results, i.e. the form factors, decay rates,
and the LPAs for the
rare $B_s\to (K,\eta,\eta')(\ell^+\ell^-,\nu_{\ell}\bar{\nu}_\ell)$
decays are presented. Summary and discussion of our main results follow in Sec. 5.

\section{Effective Hamiltonian}
\begin{figure}
\begin{center}
\includegraphics[width=3.5in,height=1.5in]{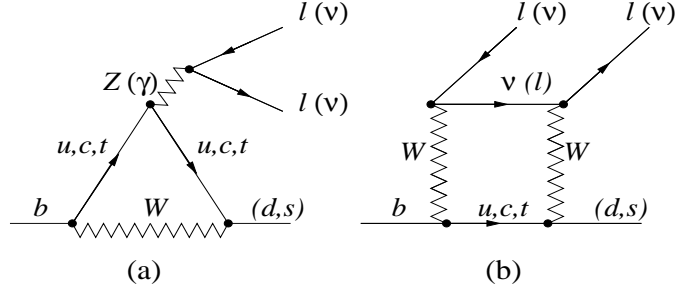}
\end{center}
\caption{ Loop diagrams for $B_s\to P_q(\ell^+\ell^-,\nu_\ell\bar{\nu}_\ell)(q=d,s)$
transitions. \label{fig1} }
\end{figure}

In the SM, the exclusive rare $B_s\to P_q(\ell^+\ell^-, \nu_\ell\bar{\nu}_\ell)$
($q=d,s$)
decays are at the quark level described by the loop
$b\to q\;(\ell^+\ell^-, \nu_\ell\bar{\nu}_\ell)$  transitions, and receive
contributions from the $Z(\gamma)$-penguin and $W$-box diagrams as shown in Fig.~\ref{fig1}.

The effective Hamiltonian responsible for the $b\to q\ell^+\ell^-\;(q=d,s)$ decay processes
can be represented in terms of the Wilson coefficients, $C_7^{\rm eff},C_9^{\rm eff}$
and $C_{10}$ as~\cite{BM}
\bea\label{Hll}
{\cal H}^{\ell^+\ell^-}_{\rm eff}&=&
\frac{G_F\al_{\rm em}}{2\sqrt{2}\pi}V_{tb}V^*_{tq}
\biggl[C^{\rm eff}_9\bar{q}\gamma_\mu (1-\gamma_5) b\bar{\ell}\gamma^\mu\ell
+ C_{10}\bar{q}\gamma_\mu (1-\gamma_5) b\bar{\ell}\gamma^\mu\gamma_5\ell
\nonumber\\
&&-C_7^{\rm eff}\frac{2m_b}{q^2}\bar{q}i\sigma_{\mu\nu}q^\nu(1+\gamma_5) b
\bar{\ell}\gamma^\mu\ell
\biggr],
\eea
where $G_F$ is the Fermi constant, $\alpha_{\rm em}$ is the fine structure
constant, and $V_{ij}$ are the Cabibbo-Kobayashi-Maskawa (CKM) matrix elements. The relevant Wilson
coefficients $C_i$ can be found in Ref.~\cite{BM}.
The effective Hamiltonian responsible for the $b\to q\nu_{\ell}\bar{\nu}_{\ell}\;(q=d,s)$
decay processes is given by~\cite{TI,BBL}
\bea\label{Hnn}
{\cal H}^{\nu_{\ell}\bar{\nu}_{\ell}}_{\rm eff}&=&
\frac{G_F\al_{\rm em}}{2\sqrt{2}\pi}V_{tb}V^*_{tq}\frac{X(x_t)}{\sin^2\theta_W}
\bar{q}\gamma_\mu (1-\gamma_5) b
\bar{\nu}_{\ell}\gamma^\mu(1-\gamma_5)\nu_{\ell},
\eea
where
$x_t=(m_t/M_W)^2$ and $X(x_t)$ is the top quark loop function~\cite{TI,BBL}, which
is given by
\be\label{ILF}
X(x) = \frac{x}{8}\biggl( \frac{2+x}{x-1} + \frac{3x-6}{(x-1)^2}\ln x\biggr).
\ee
Besides the short distance (SD) contributions, the main effect on the decay comes
from the long distance (LD) contributions due to the $c\bar{c}$ resonance states
($J/\psi,\psi',\cdots$). The effective Wilson coefficient
$C^{\rm eff}_9$ taking into account both the SD and LD contributions
has the following form~\cite{BM}
\be
C^{\rm eff}_9(s) = C_9 + Y_{SD}(s) + Y_{LD}(s),
\ee
where the explicit forms of $Y_{SD}(s)$ and $Y_{LD}(s)$ can be found in~\cite{BM,Faessler}.
For the LD contribution $Y_{LD}(s)$, we include two $c\bar{c}$ resonant states
$J/\psi(1S)$ and $\psi'(2S)$ and use $\Gamma(J/\psi\to\ell^+\ell^-)=5.26\times 10^{-6}$
GeV, $M_{J/\psi}=3.1$ GeV, $\Gamma_{J/\psi}=87\times 10^{-6}$ GeV
for $J/\psi(1S)$ and $\Gamma(\psi'\to\ell^+\ell^-)=2.12\times 10^{-6}$
GeV, $M_{\psi'}=3.69$ GeV, $\Gamma_{\psi'}=277\times 10^{-6}$ GeV
for $\psi'(2S)$~\cite{Data08}.

The LD contributions to the exclusive $B_s\to P_q\;(q=d,s)$ decays are
contained in the meson matrix elements of the bilinear quark currents
appearing in ${\cal H}^{\ell^+\ell^-}_{\rm eff}$ and
${\cal H}^{\nu_{\ell}\bar{\nu_{\ell}}}_{\rm eff}$.
In the matrix elements
of the hadronic currents for $B_s\to P_q$ transitions, the parts containing
$\gamma_5$ do not contribute. Considering Lorentz and parity invariances,
these matrix elements can be parametrized in terms of hadronic form factors
as follows:
\be\label{Jmu}
J^\mu\equiv\la P_q|\bar{q}\gamma^{\mu} b|B_s\ra=
f_{+}(q^{2})P^\mu + f_{-}(q^{2})q^\mu,
\ee
and
\bea\label{JTmu}
J^\mu_T&\equiv&\la P_q|\bar{q}i\sigma^{\mu\nu}q_\nu b|B_s\ra
= \frac{f_T(q^2)}{M_{B_s}+M_{P_q}}[q^2 P^\mu - (M^2_{B_s}-M^2_{P_q})q^\mu],
\eea
where $P=P_{B_s}+P_{P_q}$ and $q=P_{B_s}-P_{P_q}$ is the four-momentum
transfer to the lepton pair and $4m^{2}_{\ell}\leq q^{2}\leq(M_{B_s}-M_{P_q})^2$.
We use the convention $\sigma^{\mu\nu}=(i/2)[\gamma^\mu,\gamma^\nu]$ for
the antisymmetric tensor.
Sometimes it is useful to express Eq.~(\ref{Jmu}) in terms
of $f_+(q^2)$ and $f_0(q^2)$, which are related to the exchange
of $1^-$ and $0^+$, respectively, and satisfy the following relations:
\be\label{F0}
f_+(0)=f_0(0),\;
f_0(q^2)=f_+(q^2) + \frac{q^2}{M^2_{B_s}-M^2_{P_q}}f_-(q^2).
\ee
With the help of the effective Hamiltonian in Eq.~(\ref{Hll}) and
Eqs.~(\ref{Jmu}) and~(\ref{JTmu}),  the transition amplitude
${\cal M}= \la P_q\ell^+\ell^-|{\cal H}_{\rm eff}|B_s\ra$
for the $B_s\to P_q\ell^+\ell^-$ decay can be written as
\bea\label{TranA}
{\cal M}
&=&\frac{G_F\al_{\rm em}}{2\sqrt{2}\pi}V_{tb}V^*_{tq}
\biggl\{\biggl[C^{\rm eff}_9 J_\mu - \frac{2m_b}{q^2}C^{\rm eff}_7 J^T_\mu\biggr]
\bar{\ell}\gamma^\mu\ell
+ C_{10}J_\mu \bar{\ell}\gamma^\mu\gamma_5\ell \biggr\}.
\eea
The differential decay rate for $B_s\to P_q\ell^+\ell^-$
is given by~\cite{MN,GK}
\bea\label{DDR}
\frac{d\Gamma_{\ell\ell}}{ds}
&=&\frac{M^5_{B_s}G^2_F}{3\cdot2^9\pi^5}\alpha^2_{\rm em}
|V_{tb}V^*_{tq}|^2\phi_H^{1/2}
\biggl(1-\frac{4t}{s}\biggr)^{1/2}
\biggl[\phi_H\biggl(1+\frac{2t}{s}\biggr)
{\cal F}_{1} + 12t{\cal F}_{2} \biggr],
\eea
where
\bea\label{DDR2}
{\cal F}_{1} &=&
\biggl|C^{\rm eff}_9 f_+
- \frac{2\hat{m_b}C^{\rm eff}_7}{1+\sqrt{r}}f_T \biggr|^2
+ |C_{10}f_+|^2,\nonumber\\
{\cal F}_{2} &=&
|C_{10}|^2 \biggl[(1+r -\frac{s}{2})|f_+|^2
+ (1-r)f_+f_-
+\frac{s}{2}|f_-|^2\biggr],
\nonumber\\
\phi_H  &=& (s-1-r)^2-4r,
\eea
with $s=q^2/M^2_{B_s}$, $t=m^2_\ell/M^2_{B_s}$,
$\hat{m_b}=m_b/M_{B_s}$ and $r=M^2_{P_q}/M^2_{B_s}$.
Equation~(\ref{DDR}) may be written in
terms of ($f_+,f_0,f_T$) instead of ($f_+,f_-,f_T$) as discussed
in~\cite{CJK}. Note also from Eqs.~(\ref{DDR}) and~(\ref{DDR2}) that
the form factor $f_-(q^2)$ does not contribute in the massless lepton limit.

The differential decay rate for $B_s\to P_q\nu_{\ell}\bar{\nu}_{\ell}$ can be easily
obtained from the corresponding formula Eq.~(\ref{DDR})
for $B_s\to P_q\ell^+\ell^-$ by the replacement
$\hat{m}_\ell\to 0$, $C_7^{\rm eff}\to 0$, and $C^{\rm eff}_9=-C_{10}\to X(x_t)/\sin^2\theta_W$,
i.e.
\bea\label{DDRn}
\frac{d}{ds}\sum_{\ell}\Gamma_{\nu_\ell{\bar\nu}_\ell}
&=&3\frac{M^5_{B_s}G^2_F}{3\cdot2^8\pi^5\sin^4\theta_{W}}\alpha^2_{\rm em}
|V_{tb}V^*_{tq}|^2\phi_H^{3/2}|X(x_t)|^2|f_+|^2,
\eea
where the factor of 3 in the numerator corresponds to the sum over the three neutrino
flavors. Dividing Eqs.~(\ref{DDR}) and~(\ref{DDRn}) by the total width
of the $B_s$ meson, one can obtain the differential branching ratio
$d{\rm BR}(B_s\to P_q\ell^+\ell^-)/ds
=(d\Gamma(B_s\to P_q\ell^+\ell^-)/\Gamma_{\rm tot})/ds$.
As another interesting observable, the LPA, is defined as
\be\label{LPA}
P_L(s)=\frac{d\Gamma_{h=-1}/ds-d\Gamma_{h=1}/ds}
{d\Gamma_{h=-1}/ds +d\Gamma_{h=1}/ds},
\ee
where $h=+1(-1)$ denotes right (left) handed $\ell^-$ in the final state.
From Eq.~(\ref{DDR}), one obtains for $B_s\to P_q\ell^+\ell^-$
\be\label{LPA_Bc}
P_L(s)=\frac{
2\biggl(1-4\frac{t}{s}\biggr)^{1/2}
\phi_H C_{10}f_{+}
\biggl[f_+ {\rm Re}C^{\rm eff}_9 -
\frac{2\hat{m_b}C^{\rm eff}_7}{1+\sqrt{r}}f_T\biggr] }
{ \biggl[\phi_H\biggl(1+2\frac{t}{s}\biggr){\cal F}_{1}
+ 12t{\cal F}_{2} \biggr] }.
\ee
Because of the experimental difficulties of studying the polarizations
of each lepton depending on $s$ and the Wilson coefficients, it would be
better to eliminate the dependence of the LPA on $s$, by considering the
averaged form over the entire kinematical region.
The averaged LPA is defined by
\bea\label{LPA_av}
\la P_L\ra = \frac{\int^{(1-\sqrt{r})^2}_{4t} P_L\frac{dBR}{ds}ds}
{\int^{(1-\sqrt{r})^2}_{4t} \frac{dBR}{ds}ds}.
\eea

\section{Review of our LFQM}
 The key idea in our
LFQM~\cite{CJ1,CJ_PLB1,CJBc} for the ground state mesons is to treat the
radial wave function as a trial function for the variational
principle to the QCD-motivated effective Hamiltonian saturating
the Fock state expansion by the constituent quark and antiquark.
The QCD-motivated effective Hamiltonian for a description of the ground
state meson mass spectra is given by
 \be\label{Ham}
H_{q\bar{q}}= H_0 + V_{q\bar{q}}= \sqrt{m^2_q+{\vec k}^2}+\sqrt{m^2_{\bar{q}}+{\vec
k}^2}+V_{q\bar{q}},
 \ee
 where
 \be\label{pot}
 V_{q\bar{q}}=V_0 + V_{\rm hyp} = a + br^n-\frac{4\al_s}{3r} +\frac{2}{3}\frac{{\bf S}_q\cdot{\bf
S}_{\bar{q}}}{m_qm_{\bar{q}}} \nabla^2V_{\rm coul}.
 \ee
In this work, we use  the Coulomb plus linear
confining (i.e. $n=1$) potential  together with the hyperfine
interaction $\la{\bf S}_q\cdot{\bf S}_{\bar{q}}\ra=1/4\;(-3/4)$ for
the vector (pseudoscalar) meson, which enables us to analyze the
meson mass spectra and various wave-function-related observables,
such as decay constants, electromagnetic form factors of mesons in a spacelike
region, and the weak form factors for the exclusive semileptonic
and rare decays of pseudoscalar mesons in the timelike
region~\cite{CJK,CJ1,CJ_PLB1,JC_E,CJBc,Bcrare,CJNRD,ChoiRD,Choi08}.

The momentum-space LF wave function of the ground state
pseudoscalar  mesons is given by $\Psi(x_i,{\bf
k}_{i\perp},\lam_i) ={\cal R}_{\lam_1\lam_2}(x_i,{\bf k}_{i\perp})
\phi(x_i,{\bf k}_{i\perp})$, where $\phi(x_i,{\bf k}_{i\perp})$ is
the radial wave function and ${\cal R}_{\lam_1\lam_2}$ is the
covariant spin-orbit wave function. The model wave function is represented by the
Lorentz-invariant variables, $x_i=p^+_i/P^+$, ${\bf
k}_{i\perp}={\bf p}_{i\perp}-x_i{\bf P}_\perp$ and $\lam_i$, where
$P^\mu=(P^+,P^-,{\bf P}_\perp) =(P^0+P^3,(M^2+{\bf
P}^2_\perp)/P^+,{\bf P}_\perp)$ is the momentum of the meson $M$,
and $p^\mu_i$ and $\lam_i$ are the momenta and the helicities of
constituent quarks, respectively.

The covariant form of the spin-orbit wave function
for pseudoscalar mesons is given by
 \bea\label{R00_A}
{\cal R}_{\lam_1\lam_2}
&=&\frac{-\bar{u}_{\lam_1}(p_1)\gamma_5 v_{\lam_2}(p_2)}
{\sqrt{2}\sqrt{M^2_0-(m_1-m_2)^2}},
 \eea
where $M^2_0=\sum_{i=1}^2({\bf k}^2_{i\perp}+m^2_i)/x_i$ is
the boost invariant meson mass square obtained from the free
energies of the constituents in mesons.
For the radial wave function $\phi$, we use the
Gaussian wave function:
\be\label{rad}
 \phi(x_i,{\bf
k}_{i\perp})=\frac{4\pi^{3/4}}{\beta^{3/2}} \sqrt{\frac{\partial
k_z}{\partial x}} {\rm exp}(-{\vec k}^2/2\beta^2),
 \ee
 where $\beta$ is the variational parameter.
  When the longitudinal
component $k_z$ is defined by $k_z=(x-1/2)M_0 +
(m^2_2-m^2_1)/2M_0$, the Jacobian of the variable transformation
$\{x,{\bf k}_\perp\}\to {\vec k}=({\bf k}_\perp, k_z)$ is given by
 \be\label{jacob} 
 \frac{\partial k_z}{\partial
x}=\frac{M_0}{4x_1x_2} \biggl\{ 1-
\biggl[\frac{m^2_1-m^2_2}{M^2_0}\biggr]^2\biggr\}.
 \ee
The normalization factor in Eq.~(\ref{rad}) is obtained from the
following normalization of the total wave function:
 \be\label{norm} 
 \int^1_0dx\int\frac{d^2{\bf k}_\perp}{16\pi^3}
|\Psi(x,{\bf k}_{i\perp})|^2=1. 
 \ee
 
We apply our
variational principle to the QCD-motivated effective Hamiltonian
first to evaluate the expectation value of the central Hamiltonian
$H_0+V_0$, i.e., $\la\phi|(H_0+V_0)|\phi\ra$, with a trial
function $\phi(x_i,{\bf k}_{i\perp})$ that depends on the
variational parameter $\beta$. Once the model
parameters are fixed by minimizing the expectation value
$\la\phi|(H_0+V_0)|\phi\ra$, the mass eigenvalue of each meson is
obtained as $M_{q\bar{q}}=\la\phi|(H_0+V_{q\bar{q}})|\phi\ra$.
 Minimizing energies with
respect to $\beta$ and searching for a fit to the observed ground
state meson spectra, our central potential $V_0$ obtained from our
optimized potential parameters ($a=-0.72$ GeV, $b=0.18$ GeV$^2$,
and $\al_s=0.31$)~\cite{CJ1} for the Coulomb plus linear potential
was found to be quite comparable with the quark potential model
suggested by Scora and Isgur~\cite{SI}, where they obtained
$a=-0.81$ GeV, $b=0.18$ GeV$^2$, and $\al_s=0.3\sim 0.6$ for the
Coulomb plus linear confining potential. A more detailed procedure
for determining the model parameters of light- and heavy-quark
sectors can be found in our previous works~\cite{CJ1,CJ_PLB1}.
Our model parameters $(m_q,\beta_{q\bar{q}})$  obtained from the linear
potential model relevant to this work are summarized in Table~\ref{t1n}. The
predictions of the ground state meson mass spectra can be found in~\cite{CJBc}.
We should note that our model parameters ($m, \beta$) automatically satisfies 
the normalization of the
total wave function and were fixed by the variational principle to 
the QCD-motivated effective Hamiltonian. 
Those parameters in turn automatically satisfies the normalization of the 
electromagnetic form factors at $q^2=0$ and every other 
physical observables obtained from our LFQM such as decay constants and 
electroweak form factors  
are the predictions. This distinguishes our LFQM from other quark model.

\begin{table}[t]
\caption{The constituent quark masses[GeV] and the Gaussian
parameters $\beta$[GeV] obtained from  the linear potential in~\cite{CJ_PLB1},
which are necessary for $B_s\to (K,\eta,\eta')$ decay modes.
$q=u$ and $d$.}\label{t1n}
\begin{tabular*}{\textwidth}{@{}l*{15}{@{\extracolsep{0pt plus
12pt}}l}}
\br
$m_q$ & $m_s$ &  $m_b$ & $\beta_{qs}$ & $\beta_{ss}$ & $\beta_{sb}$ \\
\mr
0.22 &  0.45 & 5.2 & 0.3886 & 0.4128 & 0.5712 \\
\br
\end{tabular*}
\end{table}

\subsection{Form factors for the rare $B_s\to P$ decays in our LFQM}

Most popular phenomenological LFQM uses the Gaussian wave function as the
radial wave function due to its predictive power of various physical observables.
However, since the LFQM using the Gaussian wave function
does not have a counterpart of manifestly covariant model,
it is hard to check the covariance of the model. 
To check the covariance of LFQM, one can start from the manifestly covariant field theory model.
For example, using the exactly solvable covariant BS model of (3+1)-dimensional fermion field
theory~\cite{BCJ01,MF,Jaus99}, one can perform the
LF calculation in parallel with the manifestly covariant calculation and compare the results 
from the two models. 
Comparing the LF results and the manifestly covariant results, 
we were able to derive the LF covariant form factors 
between two pseudoscalar meson explicitly and to analyze the zero-mode complication.
Since the detailed procedure of finding LF covariant transition form 
factors ($f_+,f_-,f_T$) was already given in our previous 
works~\cite{CJBc,Bcrare}, we shall briefly describe the essential procedure 
of obtaining the LF covariant form factors 
from the exactly solvable covariant BS model of (3+1)-dimensional fermion field 
theory and show the results of the LF covariant form factors.

\begin{figure}
\begin{center}
\includegraphics[width=4.5in,height=1.5in]{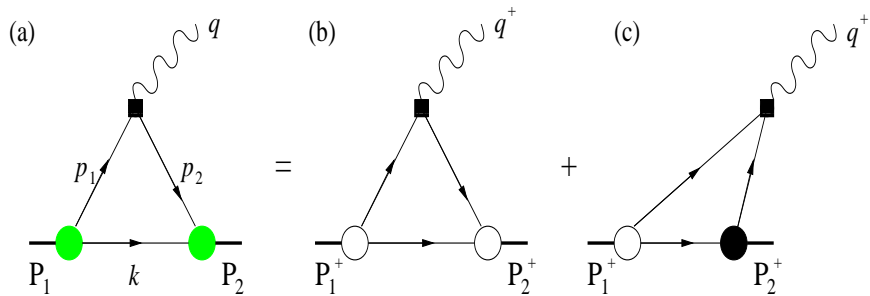}
\end{center}
\caption{ The covariant diagram (a) corresponds to the sum of the
LF valence diagram (b) defined in $0<k^+<P^+_2$ region and the
nonvalence diagram (c) defined in $P^+_2<k^+<P^+_1$ region. The large white
and black blobs at the meson-quark vertices in (b) and (c) represent
the ordinary LF wave functions and the non-wave-function vertex,
respectively. The small black box at the quark-gauge boson vertex
indicates the insertion of the relevant Wilson operator. \label{fig2} }
\end{figure}

 The covariant diagram shown in
Fig.~\ref{fig2}(a) is in general equivalent to the sum of the LF valence diagram 2(b) and
the nonvalence diagram 2(c). The matrix element
$J^\mu_{(T)}$ obtained from the covariant diagram of Fig.~\ref{fig2}(a) is given by
 \be\label{ap:3n}
 J^\mu_{(T)} = ig_1 g_2
\Lambda^2_1\Lambda^2_2\int\frac{d^4k}{(2\pi)^4} \frac{S^\mu_{(T)}} {N_{\Lambda_1}
N_{1} N_{k} N_{2} N_{\Lambda_2}},
 \ee
where $g_1$ and $g_2$ are the normalization factors which can be fixed by
requiring both charge form factors of pseudoscalar mesons to be unity at zero
momentum transfer, respectively.  To regularize the covariant fermion
triangle-loop in (3+1) dimensions, we replace the point gauge-boson vertex
$\gamma^\mu$ by a non-local smearing gauge-boson vertex
$({\Lambda_1}^2 / N_{\Lambda_1})\gamma^\mu
( {\Lambda_2}^2 / N_{\Lambda_2})$, where $N_{\Lambda_1}
=p_1^2-{\Lambda_1}^2+i\ep$ and $N_{\Lambda_2}
=p_2^2-{\Lambda_2}^2+i\ep$, and thus the factor
$({\Lambda_1}{\Lambda_2})^2$ appears in the normalization factor. $\Lambda_1$
and $\Lambda_2$ play the role of momentum cut-offs similar to the Pauli-Villars
regularization~\cite{BCJ01}. Our replacement of $\gamma^\mu$ by the non-local
smearing gauge-boson vertex remedies the conceptual difficulty associated with the
asymmetry appearing if the fermion loop were regulated by smearing the $q\bar{q}$
bound-state vertex. The rest of the denominators in Eq.~(\ref{ap:3n}),
i.e., $N_{1} N_{k} N_{2}$, are coming from the intermediate fermion
propagators in the triangle loop diagram and are given by
\be\label{ap:4}
N_{k} = k^2 - m^2 + i\ep, \; N_{j} = p_j^2 -{m_j}^2 + i\ep \;(j=1,2),
 \ee
where $m_1$, $m$, and  $m_2$ are the masses of the constituents carrying the intermediate
four-momenta $p_1=P_1 -k$, $k$, and $p_2=P_2 -k$, respectively.
Furthermore, the trace terms  $S^\mu$ from the vector current
and $S^\mu_T$ from the tensor current  are given by
 \bea\label{ap:5}
 S^\mu &=& {\rm Tr}[\gamma_5(\not\!p_1 + m_1)\gamma^\mu (\not\!p_2
+m_2)\gamma_5(-\not\!k + m)]
 \nonumber\\
 &=&4 \{ p^\mu_1(p_2\cdot k + m_2m) + p^\mu_2 (p_1\cdot k + m_1m)
 + k^\mu(m_1m_2 - p_1\cdot p_2)\},
 \eea
and
\bea\label{ap:5T}
 S^\mu_T &=& {\rm Tr}[\gamma_5(\not\!p_1 + m_1)i\sigma^{\mu\nu}q_\nu (\not\!p_2
+m_2)\gamma_5(-\not\!k + m)]
 \nonumber\\
 &=&-4 \{ p^\mu_1 [m(p_2\cdot q) + m_2(k\cdot q)] - p^\mu_2[ m(p_1\cdot q)  + m_1(k\cdot q)]
\nonumber\\
&&
 + k^\mu [m_1(p_2\cdot q) - m_2(p_1\cdot q)] \},
 \eea
respectively. By doing the integration over $k^-$ in Eq.~(\ref{ap:3n}), one finds the two LF time-ordered
contributions to the residue calculations corresponding to the two poles in $k^-$,
the LF valence contribution [Fig.~\ref{fig2}(b)] defined in $0<k^+<P^+_2$ region and the
nonvalence contribution [Fig.~\ref{fig2}(c)] defined in $P^+_2<k^+<P^+_1$ region.
The nonvalence contribution [Fig.~\ref{fig2}(c)]
in the $q^+>0$ frame corresponds to the zero mode (if it exists) in the $q^+\to 0$ limit~\cite{Zero}.
Performing the LF calculation of Eq.~(\ref{ap:3n}) in the $q^+=0$ frame in parallel with
the manifestly covariant calculation, we use the plus component of the currents
to obtain the form factors $f_+(q^2)$ and $f_T(q^2)$. For the form factor $f_-(q^2)$,
we use both the plus and perpendicular components of the currents.
As we have shown in~\cite{CJBc,Bcrare}, while the form factors
$f_+(q^2)$ and $f_T(q^2)$ can be obtained only from the valence contribution in the
$q^{+}= 0$ frame without encountering the zero-mode contribution, the form
factor $f_-(q^2)$ receives the zero mode.
In our recent analysis of semileptonic $B_c$ decays~\cite{CJBc}, we
identified the zero-mode operator that is convoluted with the initial and final
state LF valence wave functions to generate the zero-mode contribution to the form
factor $f_-(q^2)$ in the $q^+=0$ frame.
Our method can also be realized effectively by the method presented
by Jaus~\cite{Jaus99} using the orientation of the LF plane characterized by the
invariant equation $\omega\cdot x=0$, where $\omega$ is an arbitrary light-like four
vector. More detailed analysis of the zero-mode operator and
the LF covariance of the form factors $f_{\pm}$ and $f_T$ can be found in~\cite{CJBc,Bcrare}.
While the manifestly covariant BS model of fermion field theory model is good for the
qualitative analysis of the exclusive rare decays, it is still semi-realistic.
We thus replace the LF vertex functions in the BS model with the
more phenomenological Gaussian radial wave
functions in our LFQM since the zero-mode operator is independent from
the choice of radial wave function as discussed in~\cite{CJBc}.

The LF covariant form factors $f_\pm(q^2)$ and $f_T(q^2)$ for
$B_s(q_1\bar{q})\to P(q_2\bar{q})$ transitions
obtained from the $q^+=0$ frame are given by (see
~\cite{CJBc, Bcrare} for more detailed derivations)

\be\label{fp}
 f_{+}(q^2) = \int^{1}_{0}dx\int
\frac{d^{2}{\bf k}_{\perp}}{16\pi^3}
\frac{\phi_{1}(x,{\bf k}_{\perp})}{\sqrt{ {\cal A}_{1}^{2}
 + {\bf k}^{2}_{\perp}}}
\frac{\phi_{2}(x,{\bf k}'_{\perp})}{\sqrt{ {\cal A}_{2}^{2}
+ {\bf k}^{\prime 2}_{\perp}}}
( {\cal A}_{1}{\cal A}_{2}+{\bf k}_{\perp}\cdot{\bf k'}_{\perp} ),
 \ee
 \bea\label{fm}
 f_-(q^2) &=& \int^1_0 (1-x) dx
 \int \frac{ d^2{\bf k}_\perp } { 16\pi^3 }
  \frac{ \phi_1 (x, {\bf k}_\perp) } {\sqrt{ {\cal A}^2_1 + {\bf k}^2_\perp }}
  \frac{ \phi_2 (x, {\bf k'}_\perp) } {\sqrt{ {\cal A}^2_2 + {\bf k}^{\prime 2}_\perp }}
   \biggl\{ -x(1-x) M^2_1
 \nonumber\\
 && - {\bf k}^2_\perp - m_1m + (m_2 - m){\cal A}_1 + 2\frac{q\cdot P}{q^2} \biggl[ {\bf k}^2_\perp
 + 2\frac{ ( {\bf k}_\perp \cdot {\bf q}_\perp)^2 } {q^2} \biggr]
  \nonumber\\
  && + 2 \frac{ ( {\bf k}_\perp \cdot {\bf q}_\perp)^2 } {q^2}
  + \frac{ {\bf k}_\perp \cdot {\bf q}_\perp } {q^2}  [ M^2_2 - (1-x) (q^2 + q\cdot P) + 2 x M^2_0
  \nonumber\\
 && - (1 - 2x) M^2_1 - 2(m_1 - m) (m_1 + m_2) ] \biggr\},
 \eea
 \bea\label{fT}
f_T(q^2) &=& (M_1 + M_2) \int^1_0 (1-x) dx
 \int \frac{ d^2{\bf k}_\perp } { 16\pi^3 }
  \frac{ \phi_1 (x, {\bf k}_\perp) } {\sqrt{ {\cal A}^2_1 + {\bf k}^2_\perp }}
  \frac{ \phi_2 (x, {\bf k'}_\perp) } {\sqrt{ {\cal A}^2_2 + {\bf k}^{\prime 2}_\perp }}
  \nonumber\\
 &&\times \biggl[ {\cal A}_1 - (m_1 - m_2) \frac{{\bf k}_\perp\cdot{\bf q}_\perp}{ q^2}
             \biggr],
  \eea
where ${\bf k'}_\perp={\bf k}_\perp + (1-x){\bf q}_\perp$,
${\cal A}_{i}= (1-x) m_{i} + x m$ ($i=1,2$), and $q\cdot P=M^2_1-M^2_2$ with
$M_1$ and $M_2$ being the physical masses of the initial and final state mesons, respectively.
Our results for the form factors given by Eqs.~(\ref{fp})-(\ref{fT}) are essentially the same
as those presented in~\cite{CCH04}.
We should note that the LF covariant form factor $f_-(q^2)$ in Eq.~(\ref{fm}) is the sum
of the valence contribution $f^{\rm val}_-(q^2)$ and the zero-mode contribution
$f^{\rm Z.M.}_-(q^2)$~\cite{CJBc}.
Since the form factors $f_{\pm}(q^2)$ and $f_{T}(q^2)$
are defined in the spacelike ($q^2=-{\bf q}^2_\perp <0$) region, we then analytically continue them to
the timelike $q^{2}>0$ region by changing ${\bf q}_{\perp}^2$ to $-q^2$ in
the form factors. We also compare our analytic solutions with the
double pole parametric form given by
\be\label{Pole}
f_i(q^2)=\frac{f_i(0)}{1-\sigma_1 s+\sigma_2 s^2},
\ee
where $\sigma_1$ and $\sigma_2$ are the fitted parameters.

\subsection{$\eta-\eta'$ mixing for the $B_s\to \eta^{(\prime)}$ decays }
In this subsection, we discuss the $\eta-\eta'$ mixing to obtain the
$B_s\to\eta^{(\prime)}$ transition form factors.
The octet-singlet mixing angle $\theta$ of $\eta$ and $\eta'$
is known to be in the range of $-10^{\rm o}$ to $-23^{\rm o}$~\cite{Data08}.
The physical $\eta$ and $\eta'$ are the mixtures of the flavor $SU(3)$ octet
$\eta_8$ and singlet $\eta_0$ states:
 \be\label{eet}
 \left( \begin{array}{cc}
 \eta\\
 \eta'
 \end{array}\,\right)
 =\left( \begin{array}{cc}
 \cos\theta\;\; -\sin\theta\\
 \sin\theta\;\;\;\;\;\cos\theta
 \end{array}\,\right)\left( \begin{array}{c}
 \eta_8\\
 \eta_0
 \end{array}\,\right),
 \ee
 where
$\eta_8=(u\bar{u}+d\bar{d}-2s\bar{s})/\sqrt{6}$ and
$\eta_0=(u\bar{u}+d\bar{d} + s\bar{s})/\sqrt{3}$. Analogously,
in terms of the quark-flavor(QF) basis $\eta_q=(u\bar{u}+d\bar{d})/\sqrt{2}$ and
 $\eta_s=s\bar{s}$, one obtains~\cite{FKS}
  \be\label{eea}
 \left( \begin{array}{cc}
 \eta\\
 \eta'
 \end{array}\,\right)
 =\left( \begin{array}{cc}
 \cos\phi\;\; -\sin\phi\\
 \sin\phi\;\;\;\;\;\cos\phi
 \end{array}\,\right)\left( \begin{array}{c}
 \eta_q\\
 \eta_s
 \end{array}\,\right).
 \ee
 The two schemes are equivalent to each
 other by $\phi=\theta+ \arctan\sqrt{2}$ when ${\rm SU}_f(3)$ symmetry is perfect.
Although it was
frequently assumed that the decay constants follow the same
pattern of state mixing, the mixing properties of
the decay constants will generally be different from those of the meson state since the
decay constants only probe
the short-distance properties of the valence Fock states while the
state mixing refers to the mixing of the overall wave
function~\cite{FKS}.

Defining
$\la P(p)|J^{q(s)}_{\mu5}|0\ra = -if^{q(s)}_P p^\mu$ ($P=\eta,\eta'$)
in the QF basis, the
four parameters $f^{q}_P$ and $f^{s}_P$ can be expressed in terms of two
mixing angles ($\phi_q$ and $\phi_s$)
and two decay constants ($f_q$ and $f_s$), i.e.~\cite{FKS},
 \be\label{fqfs}
 \left( \begin{array}{cc}
 f^q_\eta  \;\;\;\;\; f^s_\eta\\
 f^q_{\eta'} \;\;\;\;\; f^s_{\eta'}
 \end{array}\,\right)
 = \left( \begin{array}{cc}
 \cos\phi_q\;\; -\sin\phi_s\\
 \sin\phi_q\;\;\;\;\;\cos\phi_s
 \end{array}\,\right)\left( \begin{array}{cc}
 f_q\;\; 0\\
 0\;\;\;f_s
 \end{array}\,\right).
 \ee
The difference between the mixing angles $\phi_q-\phi_s$ is due to the
Okubo-Zweig-Iizuka(OZI)-violating effects~\cite{OZI} and is found to be
small ($\phi_q-\phi_s<5^{\circ}$).
The OZI rule implies that the difference between
$\phi_q$ and $\phi_s$ vanishes (i.e., $\phi_q=\phi_s=\phi$)
to leading order in the $1/N_c$ expansion. Similarly, the four parameters
$f^{8}_P$ and $f^{0}_P$ in the octet-singlet basis
may be written in terms of two angles ($\theta_8$ and $\theta_0$)
and two decay constants ($f_8$ and $f_0$). However, in this
case, $\theta_8$ and $\theta_0$ turn out to differ
considerably and become equal only in the ${\rm SU}_f(3)$ symmetry limit~\cite{FKS,Leut98}.

We shall use the QF basis with the single mixing angle $\phi$ to analyze the
$B_s\to\eta^{(\prime)}$ decay modes. In this case, a generic form factor $F$ and the
branching ratio for the $B_s\to\eta^{(\prime)}$ are given by
\be\label{Fee}
F^{B_s\to\eta(\eta')} = -\sin\phi (\cos\phi) F^{B_s\to\eta_s},
\ee
with the physical $\eta^{(\prime)}$ mass and
\be\label{Gee}
{\rm BR}[B_s\to\eta(\eta')\ell^+\ell^-] = \sin^2\phi (\cos^2\phi) {\rm BR}[B_s\to\eta_s\ell^+\ell^-],
\ee
respectively. Recently, the KLOE Collaboration~\cite{KLOE} extracted the pseudoscalar mixing
angle $\phi$ in the QF basis
by measuring the ratio ${\rm BR}(\phi\to\eta'\gamma)/{\rm BR}(\phi\to\eta\gamma)$.
The measured values are $\phi=(39.7\pm 0.7)^{\circ}$ and
$(41.5\pm 0.3_{\rm stat}\pm 0.7_{\rm syst}\pm 0.6_{\rm th})^{\circ}$
with and without the gluonium content for $\eta'$, respectively.
However, since the mixing angle for $\eta-\eta'$ is still a controversial issue, we
use unspecified value for $\phi$ rather than adopting some specific value.

\section{Numerical results}

In our numerical calculations for the exclusive rare
$B_s\to (K, \eta,\eta')(\nu_{\ell}\bar{\nu_{\ell}}, \ell^+\ell^-)$ decays, we use
the model parameters ($m_q,\beta$) for the linear
confining potential given in Table~\ref{t1n}.
Although our predictions~\cite{CJBc} of ground state heavy
meson masses are overall in good agreement with the experimental
values, we use the experimental meson masses~\cite{Data08} in the
computations of the decay widths to reduce possible theoretical
uncertainties.

Note that in the numerical calculations we take
$(m_c, m_b)=(1.8,5.2)$ GeV in all formulas except in the Wilson coefficient
$C^{\rm eff}_9$, where $(m_c, m_{b,\rm pole})=(1.4,4.8)$ GeV  have been commonly
used~\cite{BM}. For the numerical values of the Wilson coefficients,
we use the results given by Ref.~\cite{BM}:
\bea\label{WC}
C_1 &=&-0.248,\; C_2=1.107, \; C_3=0.011,
\nonumber\\
C_4 &=&-0.026, \;C_5=0.007, \; C_6=-0.031,
\nonumber\\
C^{\rm eff}_7 &=&-0.313,\; C_9=4.344,\; C_{10}=-4.669,
\eea
and other input parameters are $|V_{tb}V^*_{ts}|=0.039$,
$|V_{tb}V^*_{td}|=0.008$, $\al_{\rm em}^{-1}=129$, $M_W=80.43$ GeV,
$m_t=171.3$ GeV, $\sin^2\theta_W=0.2233$,
and $\tau_{B^0_s}=1.470$ ps.

\begin{figure}
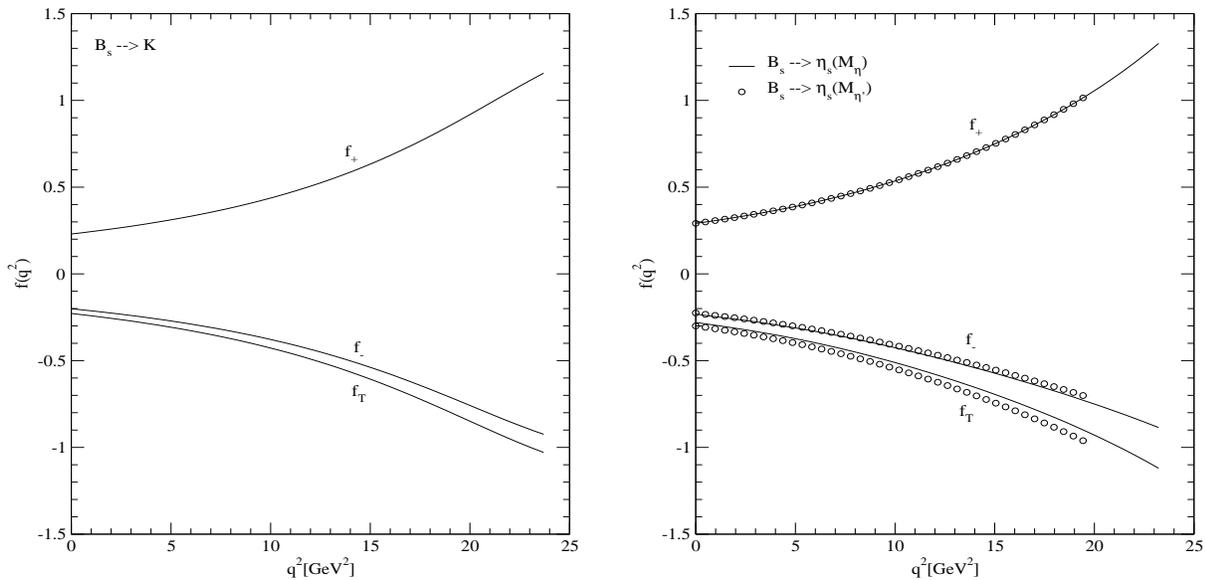

\vspace{1cm}
\includegraphics[width=3in,height=3in]{fig3a.eps}
\hspace{0.5cm}
\includegraphics[width=3in,height=3in]{fig3b.eps}
\caption{The weak form factors
 for $B_s\to K$(left panel) and $B_s\to \eta_s$(right panel) decays, respectively.}
\label{fig3}
\end{figure}
\begin{table}
\caption{Results for form factors at $q^2=0$ of $B_s\to (K,\eta,\eta')$
transition and parameters
$\sigma_i$ defined in Eq.~\protect(\ref{Pole}). The coefficients
in $\eta$ and $\eta'$ represent quark mixing angles, i.e.
 $c_\eta =-\sin\phi$ and $c_{\eta'}=\cos\phi$, respectively.}
\begin{tabular*}{\textwidth}{@{}l*{15}{@{\extracolsep{0pt plus
12pt}}l}}
\br
Mode & $f_+(0)$ & $\sigma_1$ & $\sigma_2$
      & $f_-(0)$ & $\sigma_1$ & $\sigma_2$
      & $f_T(0)$ & $\sigma_1$ & $\sigma_2$\\
\mr
$B_s\to K$ & 0.230  & $-1.650$ & 0.822 & $-0.201$& $-1.638 $ & 0.835
            & $-0.228$ & $-1.633 $ & 0.835  \\
$B_s\to\eta$ & $0.291c_\eta$  & $-1.574$ & 0.751 & $-0.231c_\eta$& $-1.582 $ & 0.825
            & $-0.280c_\eta$ & $-1.561 $ & 0.782  \\
$B_s\to\eta'$ & $0.291c_{\eta'}$  & $-1.575$ & 0.770 & $-0.225c_{\eta'}$& $-1.570 $ & 0.835
            & $-0.300c_{\eta'}$ & $-1.561 $ & 0.802  \\
\br
\end{tabular*}
\label{t2}
\end{table}

In Fig.~\ref{fig3}, we show the $q^2$ dependences of the
form factors $f_{\pm}(q^2)$ and $f_{T}(q^2)$  for
the $B_s\to K$ (left panel) and $B_s\to \eta_s$ with physical masses of $\eta$ and $\eta'$
(right panel), respectively.
The form factors at $q^2=0$ and the parameters $\sigma_i$ of the double pole form
defined in Eq.~(\ref{Pole}) are listed in Table~\ref{t2}.
The form factor $f_+(q^2)$ for the $B_s\to \eta_s$
has the same $q^2$ dependence (apart from the mixing angle $\phi$)
for both $\eta$ (solid line) and $\eta'$ (circle) since $f_+$ does
not depend on the daughter meson mass as one can see from Eq.~(\ref{fp}). On the other hand,
the form factors $f_-$ and $f_T$ between $\eta$ and $\eta'$
are slightly different (apart from the mixing angle $\phi$) since they
involve the daughter meson mass as one can see from Eqs.~(\ref{fm}) and ~(\ref{fT}).
The form factors at the zero recoil
point (i.e., $q^2=q^2_{\rm max})$ correspond to the overlap integral
of the initial and final state meson wave functions. The
maximum recoil point (i.e., $q^2=0$) corresponds to a final state
meson recoiling with the maximum three-momentum $|{\vec
P}_f|=(M^2_{B_s}-M^2_f)/2M_{B_s}$ in the rest frame of the $B_s$
meson.
As a sensitivity check of our LFQM, we find for the $B_s\to K$ transition that
our form factors are changed only about 3$\%$ as the light $d$-quark mass varies about
20$\%$. This indicates that the transition form factors for $b\to d$ decay processes
are quite stable on the variation of $d$ quark mass.
The form factors for the $B_s\to \eta_s$ have also been computed by Geng and Liu~\cite{Geng}
using the similar LFQM but only with the valence contributions in the purely longitudinal $q^+\neq 0$ frame.
Although the form factors $f_+$ and $f_T$ at the maximum recoil point obtained from~\cite{Geng}
do not receive nonvalence contributions, they receive nonvalence contributions at other nonzero
$q^2$ values. The nonvalence contributions to $f_-$ are more serious for the entire $q^2$ range including
the $q^2=0$ point. For instance, $f_-$ obtained from~\cite{Geng}(see Fig. 2 in~\cite{Geng}) shows a
sharp increasing as $q^2$ near the zero recoil point in contrast to our result. This indicates
the nonvalence contribution to $f_-$ is quite large, which in particular overestimate
the branching ratio for the $\tau$ dilepton decay mode.
Although the form factor $f_-(q^2)$ does not contribute to the branching
ratio in the massless lepton ($\ell=e$ or $\mu$) decay, it is important for
the heavy $\tau$ decay process.

\begin{figure}
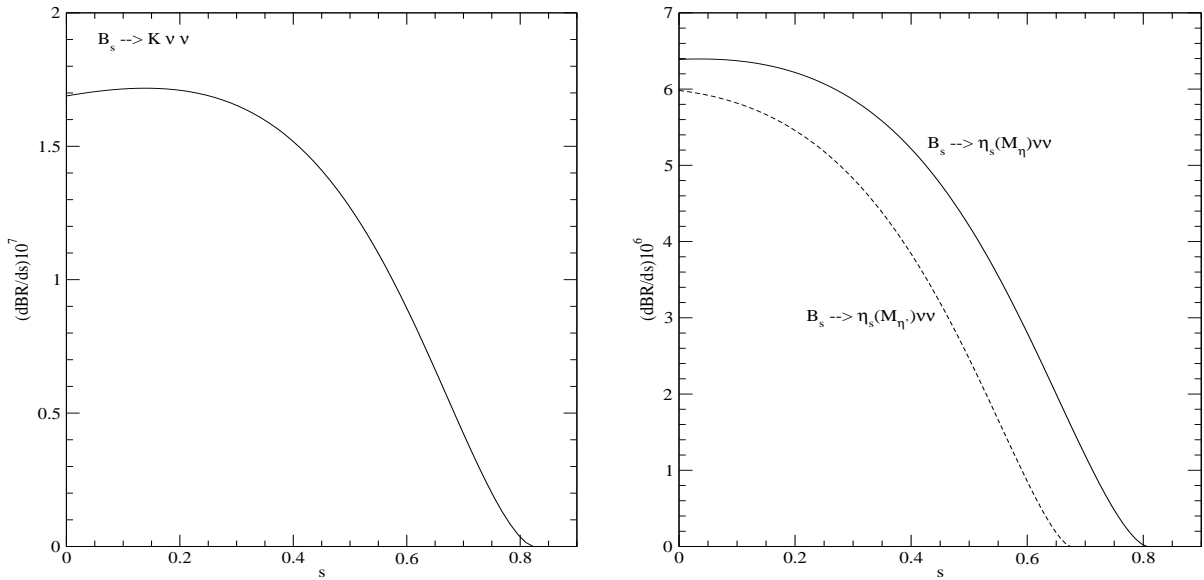

\vspace{1cm}
\includegraphics[width=3in,height=3in]{fig4a.eps}
\hspace{0.5cm}
\includegraphics[width=3in,height=3in]{fig4b.eps}
\caption{Differential branching ratios for $B_s\to K\sum\nu_{\ell}\bar{\nu_{\ell}}$ (left panel)
and $B_s\to \eta_s\sum\nu_{\ell}\bar{\nu_{\ell}}$ (right panel) decays, respectively.}
\label{fig4}
\end{figure}
\begin{figure}
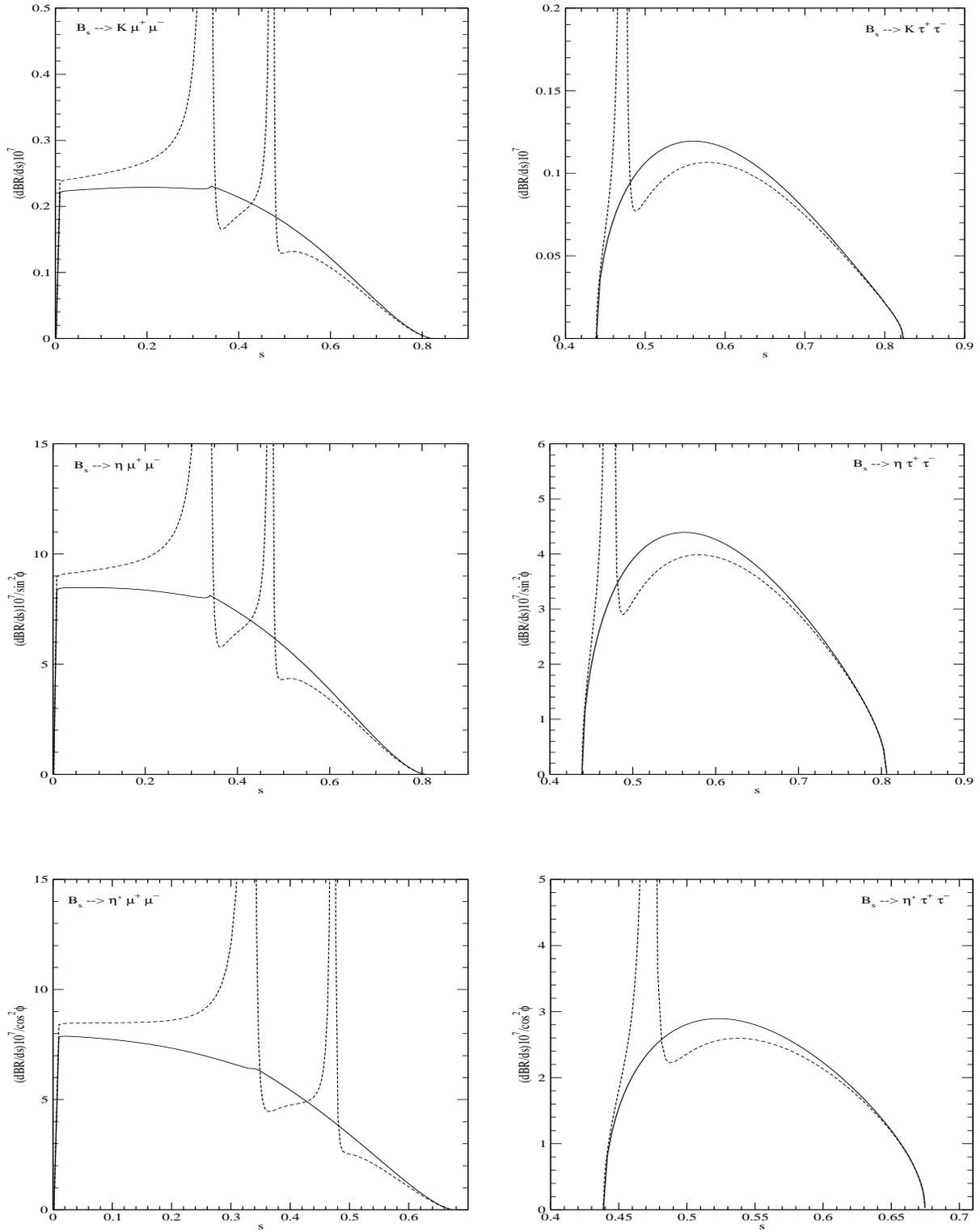

\vspace{1cm}
\includegraphics[width=3in,height=2.3in]{fig5a.eps}
\hspace{0.5cm}
\includegraphics[width=3in,height=2.3in]{fig5b.eps}
\vspace{4ex}

\includegraphics[width=3in,height=2.3in]{fig5c.eps}
\hspace{0.5cm}
\includegraphics[width=3in,height=2.3in]{fig5d.eps}
\vspace{4ex}

\includegraphics[width=3in,height=2.3in]{fig5e.eps}
\hspace{0.5cm}
\includegraphics[width=3in,height=2.3in]{fig5f.eps}
\caption{Differential branching ratios for $B_s\to K\ell^+\ell^-$ (upper
panel) and $B_s\to \eta\ell^+\ell^-$ (middle panel) and
$B_s\to \eta'\ell^+\ell^-$ (lower panel) with $\ell=\mu$ and $\tau$, respectively.
The solid and dashed lines represent the results without and with
the LD contributions, respectively.}
\label{fig5}
\end{figure}

We show our results for the differential branching ratios for
$B_s\to (K, \eta_s)\sum\nu_{\ell}\bar{\nu}_{\ell}$ with physical masses of
$\eta$ and $\eta'$ in Fig.~\ref{fig4}
and $B_s\to (K,\eta,\eta') \ell^+\ell^-$ ($\ell=\mu$ and $\tau$) in Fig.~\ref{fig5},  respectively.
For the $B_s\to (K,\eta,\eta')\ell^+\ell^-$ transitions in Fig.~\ref{fig5},
the solid (dashed) lines represent the
results without (with) the LD contribution to $C^{\rm eff}_9$.
Since the $B_s\to K$ is induced by $b\to d$ transition compared to the $B_s\to\eta_s$
induced by $b\to s$ at the quark level, the branching ratios for the final
$K$ meson are order of magnitude smaller than the corresponding branching ratios for
the final $\eta$ meson.
For the $B_s\to (K,\eta,\eta') \ell^+\ell^-$ decays (see Fig.~\ref{fig5}),
the LD contributions (dashed lines) clearly overwhelm the nonresonant branching
ratios near $J/\psi(1S)$ and $\psi'(2S)$ peaks, however, suitable
$\ell^+\ell^-$ invariant mass cuts can separate the LD contribution
from the SD one away from these peaks.
This divides the spectrum into two distinct regions~\cite{Hew,AGM}:
(i) low-dilepton mass, $4m^2_\ell\leq q^2\leq M^2_{J/\psi}-\delta$,
and (ii) high-dilepton mass,
$M^2_{\psi'}+\delta\leq q^2\leq q^2_{\rm max}$, where $\delta$ is to
be matched to an experimental cut.

Our predictions for the nonresonant branching ratios
 are summarized in Table~\ref{t3} with general form of the mixing angle $\phi$ in
 the QF basis. Our results are also
compared with other theoretical predictions such as the LF
and constituent QM~\cite{Geng} and the QCD sum rules (SR)~\cite{CCD} within
the SM. Since the amplitude $B_s\to (K,\eta,\eta')\ell^+\ell^-$ is regular at $q^2=0$, the
transitions  $B_s\to (K,\eta, \eta')e^+e^-$ and $B_s\to (K,\eta,\eta')\mu^+\mu^-$ have almost
the same decay rates, i.e. insensitive to the mass of the light lepton.
Our predictions of branching ratios are close to the QCD SR results~\cite{CCD}
but a bit smaller than the LFQM results~\cite{Geng}.
But the results from~\cite{Geng}
could be lowered if the nonvalence contributions are properly taken into account.
For the mixing angle $\theta=-20^{\circ}$ ($-10^{\circ}$) in the octet-singlet basis, which
corresponds to $\phi=34.74^{\circ}$ ($44.74^{\circ}$) in the QF basis, we obtain
${\rm BR}(B_s\to \eta\sum\nu_{\ell}\bar{\nu}_{\ell})=1.1\; (1.7)\times 10^{-6}$,
${\rm BR}(B_s\to \eta\mu^+\mu^-)=1.5 \;(2.4)\times 10^{-7}$,
${\rm BR}(B_s\to \eta\tau^+\tau^-)=3.8 \;(5.8)\times 10^{-8}$,
${\rm BR}(B_s\to \eta'\sum\nu_{\ell}\bar{\nu}_{\ell})=1.8\; (1.3)\times 10^{-6}$,
${\rm BR}(B_s\to \eta'\mu^+\mu^-)=2.4 \;(1.8)\times 10^{-7}$, and
${\rm BR}(B_s\to \eta'\tau^+\tau^-)=3.4 \;(2.6)\times 10^{-8}$, respectively.

It is also worth comparing the branching ratios between $B_s\to K$ and $B_s\to\eta$, which
may be written as
\be
\frac{{\rm BR}(B_s\to K\mu^+\mu^-)}{{\rm BR}(B_s\to \eta\mu^+\mu^-)}
= \frac{1}{\sin^2\phi}\biggl|\frac{V_{td}}{V_{ts}}\biggr|^2 (1 - \Delta_{SU(3)}),
\ee
where the $SU(3)$ correction term $\Delta_{SU(3)}$ is
estimated about $0.3$ in our model calculation. Such a kind of relation may be further
scrutinized by considering an additional correction term neglected in the effective Hamiltonian
as discussed in~\cite{CCD}.
The branching ratios with the LD contributions for $B_s\to (K,\eta,\eta')\ell^+\ell^-$ $(\ell=\mu,\tau)$
 are also presented in Table~\ref{t4}
for low- and high-dilepton mass regions of $q^2$.

\begin{table}
\caption{Nonresonant branching ratios (in units of $10^{-7}$)
for $B_s\to (K,\eta,\eta')\sum\nu_{\ell}\bar{\nu}_{\ell}$ and
$B_s\to (\eta,\eta')\ell^+\ell^-$ transitions compared with other theoretical model
predictions within the SM.}
\begin{tabular*}{\textwidth}{@{}l*{15}{@{\extracolsep{0pt plus
12pt}}l}}
\br
Mode & This work & ~\protect\cite{Geng} & ~\protect\cite{CCD} \\
\mr
$B_s\to K\sum\nu_{\ell}\bar{\nu}_{\ell}$ & 1.01 &  & \\
$B_s\to \eta\sum\nu_{\ell}\bar{\nu}_{\ell}$ & $35.1\sin^2\phi$
 & $58.3\sin^2\phi$(LFQM) & $(21.6\pm 4.6)\sin^2\phi$(set A) \\
    &  & $54.1\sin^2\phi$(CQM) & $(50.1\pm 15.9)\sin^2\phi$(set B) \\
$B_s\to \eta'\sum\nu_{\ell}\bar{\nu}_{\ell}$ & $26.2\cos^2\phi$
 & $42.1\cos^2\phi$(LFQM) & $(16.0\pm 3.6)\cos^2\phi$ (set A)  \\
   &  & $39.7\cos^2\phi$(CQM) & $(33.9\pm 8.9)\cos^2\phi$ (set B) \\
$B_s\to K\mu^+\mu^-$ & 0.14 &  & \\
$B_s\to \eta\mu^+\mu^-$ & $4.75\sin^2\phi$
& $8.53\sin^2\phi$(LFQM) & $(2.73\pm 0.68)\sin^2\phi$(set A)  \\
  &  & $7.78\sin^2\phi$(CQM) & $(5.92\pm 1.59)\sin^2\phi$ (set B) \\
$B_s\to \eta'\mu^+\mu^-$ & $3.53\cos^2\phi$
    & $6.06\cos^2\phi$(LFQM) & $(1.96\pm 0.53)\cos^2\phi$(set A) \\
   &  & $5.69\cos^2\phi$(CQM) & $(3.92\pm 1.07)\cos^2\phi$ (set B) \\
$B_s\to K\tau^+\tau^-$ & 0.03 &  & \\
$B_s\to \eta\tau^+\tau^-$ & $1.17\sin^2\phi$
& $1.67\sin^2\phi$(LFQM) & $(0.68\pm 0.11)\sin^2\phi$(set A)  \\
& & $1.67\sin^2\phi$(CQM) & $(1.82\pm 0.34)\sin^2\phi$ (set B) \\
$B_s\to \eta'\tau^+\tau^-$ & $0.51\cos^2\phi$
  & $0.83\cos^2\phi$(LFQM) & $(0.28\pm 0.05)\cos^2\phi$(set A)  \\
  & & $0.72\cos^2\phi$(CQM) & $(0.69\pm 0.13)\cos^2\phi$ (set B) \\
\br
\end{tabular*}
\label{t3}
\end{table}
\begin{table}
\caption{Branching ratios with the LD contributions
for $B_c\to (D,D_s)\ell^+\ell^-$ for low and high dilepton mass
regions of $q^2$ [GeV$^2$] obtained from the
linear (HO) potential parameters.}
\begin{tabular*}{\textwidth}{@{}l*{15}{@{\extracolsep{0pt plus
12pt}}l}}
\br
Mode & $4m^2_\ell\leq q^2\leq 8.5$&   $14.5\leq q^2\leq q^2_{\rm max}$ \\
\mr
$B_s\to K\mu^+\mu^-$  & $7.72\;(6.63)\times10^{-9}$ & $2.27 \;(2.62)\times10^{-9}$ \\
$B_s\to K\tau^+\tau^-$  &  & $2.43 \;(2.66)\times10^{-9}$ \\
$B_s\to \eta\mu^+\mu^-$  & $2.86 \;(2.44)\sin^2\phi\times10^{-7}$
&  $7.00 \;(8.10)\sin^2\phi\times10^{-8}$ \\
$B_s\to \eta\tau^+\tau^-$  &  & $9.13 \;(9.85)\sin^2\phi\times10^{-8}$ \\
$B_s\to \eta'\mu^+\mu^-$  & $2.54 \;(2.17)\cos^2\phi\times10^{-7}$
&  $2.54 \;(2.13)\cos^2\phi\times10^{-8}$\\
$B_s\to \eta'\tau^+\tau^-$  &  & $3.42 \;(3.67)\cos^2\phi\times10^{-8}$ \\
\br
\end{tabular*}
\label{t4}
\end{table}

\begin{figure}
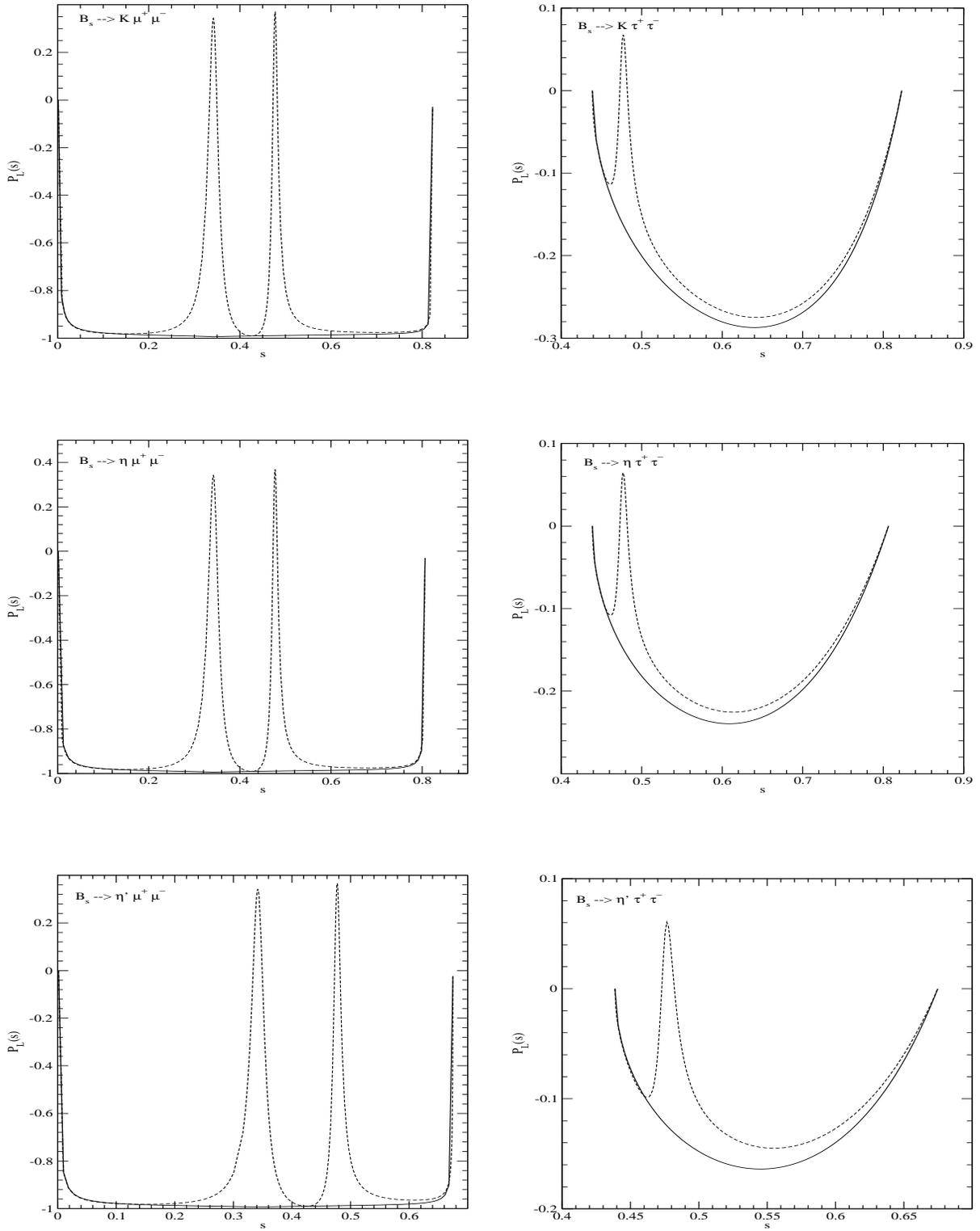

\vspace{1cm}
\includegraphics[width=3in,height=2.3in]{fig6a.eps}
\hspace{0.5cm}
\includegraphics[width=3in,height=2.3in]{fig6b.eps}
\vspace{4ex}

\includegraphics[width=3in,height=2.3in]{fig6c.eps}
\hspace{0.5cm}
\includegraphics[width=3in,height=2.3in]{fig6d.eps}
\vspace{4ex}

\includegraphics[width=3in,height=2.3in]{fig6e.eps}
\hspace{0.5cm}
\includegraphics[width=3in,height=2.3in]{fig6f.eps}
\caption{Longitudinal lepton polarization asymmetries for $B_s\to K\ell^+\ell^-$(upper
panel), $B_s\to \eta\ell^+\ell^-$(middle panel) and $B_s\to \eta'\ell^+\ell^-$(lower panel).
The same line codes are used as in Fig.~\ref{fig5}.}
\label{fig6}
\end{figure}

In Fig.~\ref{fig6}, we show the
LPAs for  $B\to (K,\eta,\eta')\ell^+\ell^-$ ($\ell=\mu,\tau$) as a function of $s$.
In both $\mu$ and $\tau$ dilepton decays, the LPAs
become zero at the end point regions of $s$. However, we note that if $m_\ell =0$, the LPAs
are not zero at the end points. As in the case of the $B\to K\mu^+\mu^-$~\cite{CJK,GK,MN,HQ}
and $B_c\to D_{(s)}\mu^+\mu^-$~\cite{Bcrare} decays where $P_L\simeq -1$
away from the end point regions, the LPAs away from the end point regions
are also close to $-1$ for the $B_s\to (K,\eta,\eta')\mu^+\mu^-$
transitions.
In fact, the $P_L$ for the muon decay is insensitive to the form factors,
e.g. our $P_L$ away from the end point regions is well
approximated by~\cite{HQ}
\be\label{PLmu}
P_L\simeq 2\frac{C_{10}{\rm Re}C^{\rm eff}_9}{|C^{\rm eff}_9|^2
+ |C_{10}|^2}\simeq -1,
\ee
in the limit of $C^{\rm eff}_7\to 0$ from Eq.~(\ref{LPA_Bc}). It also shows that
the $P_L$ for the $\mu$ dilepton channel
is insensitive to the little variation of $C^{\rm eff}_7$ as expected.
On the other hand, the LPA for the $\tau$ dilepton channel is sensitive
to the form factors. Similar observation has also been made in our recent
work for $B_c\to (D,D_s)\ell^+\ell^-$ decays~\cite{Bcrare}.

The averaged values $\la P^{K,\eta^{(\prime)}}_L\ra_\ell$
of the LPAs for $B_s\to (K,\eta^{(\prime)})\ell^+\ell^-$ without the LD contributions
are
$\la P^K_L\ra_\mu=\la P^\eta_L\ra_\mu=\la P^{\eta'}_L\ra_\mu=-0.98$, $\la P^K_L\ra_\tau=-0.24$,
 $\la P^\eta_L\ra_\tau=-0.20$ and $\la P^{\eta'}_L\ra_\tau=-0.14$,
respectively.

\section{Summary and Discussion}
In this work, we investigated the exclusive rare semileptonic
$B_s\to (K,\eta,\eta')(\nu_{\ell}\bar{\nu_{\ell}},\ell^+\ell^-)$ ($\ell=e,\mu,\tau$) decays within the SM,
using  our LFQM constrained by the variational principle for the QCD
motivated effective Hamiltonian with the linear  plus Coulomb interaction~\cite{CJ1,CJ_PLB1}.
Our model parameters obtained from the
variational principle uniquely determine the physical quantities
related to the above processes. This approach can establish the
broader applicability of our LFQM to the wider range of hadronic
phenomena.
The weak form factors $f_{\pm}(q^2)$ and $f_T(q^2)$ for the $B_s\to (K,\eta,\eta')$ decays
are obtained in the $q^+=0$ frame ($q^2=-{\bf q}^2_\perp<0$) and then
analytically continued to the timelike region by changing ${\bf
q}^2_\perp$ to $-q^2$ in the form factors. The covariance (i.e.,
frame independence) of our model has been checked by performing the
LF calculation in parallel with the manifestly
covariant calculation using the exactly solvable covariant fermion
field theory model in $(3+1)$-dimensions.
While the form factors $f_+(q^2)$ and $f_T(q^2)$ are immune to the zero modes,
the form factor $f_-(q^2)$ is not free from the zero mode.
Using the solutions of the weak form factors obtained from the $q^+=0$ frame,
we calculated the branching ratios for
$B_s\to (K,\eta,\eta')(\nu_{\ell}\bar{\nu_{\ell}},\ell^+\ell^-)$ and the LPAs
for $B_s\to (K,\eta,\eta')\ell^+\ell^-$ including both
short- and long-distance contributions from the QCD Wilson coefficients.
Our numerical results for the nonresonant branching ratios of
$B_s\to \eta^{(\prime)}(\sum\nu_{\ell}\bar{\nu_{\ell}},\mu^+\mu^-,\tau^+\tau^-)$
decays
are ${\cal O}(10^{-6},10^{-7}, 10^{-8})$ in orders of magnitude, respectively.
The branching ratios for
the $B_s\to K(\nu_{\ell}\bar{\nu_{\ell}},\ell^+\ell^-)$ decays are at least an order of
magnitude smaller than
those for the $B_s\to \eta^{(\prime)}(\nu_{\ell}\bar{\nu_{\ell}},\ell^+\ell^-)$ decays.
The averaged values $\la P^{K,\eta^{(\prime)}}_L\ra_\ell$
of the LPAs for $B_s\to (K,\eta^{(\prime)})\ell^+\ell^-$ without the LD contributions
are
$\la P^K_L\ra_\mu=\la P^\eta_L\ra_\mu=\la P^{\eta'}_L\ra_\mu=-0.98$, $\la P^K_L\ra_\tau=-0.24$,
 $\la P^\eta_L\ra_\tau=-0.20$ and $\la P^{\eta'}_L\ra_\tau=-0.14$,
respectively.
These polarization asymmetries provide valuable
information on the flavor changing loop effects in the SM.
Of particular interest, we estimated that
the ratio ${\rm BR}(B_s\to K\mu^+\mu^-)/{\rm BR}(B_s\to \eta\mu^+\mu^-)$
differs from the ${\rm SU}_f(3)$ symmetry limit (apart from the mixing angle)
by about $30\%$. Such a kind of relation may help in determining the $\eta-\eta'$
mixing angle.

\ack This work  was supported by the Korea
Research Foundation Grant funded by the Korean
Government(KRF-2008-521-C00077).

\end{document}